# High quality monolayer graphene synthesized by resistive heating cold wall chemical vapour deposition

*Thomas H. Bointon, Matthew D. Barnes, Saverio Russo, and Monica F. Craciun\**

T. H. Bointon, M. D. Barnes, Prof S. Russo, Prof M. F. Craciun,
Centre for Graphene Science, College of Engineering, Mathematics and Physical Sciences,
University of Exeter, Exeter, EX4 4QL, UK
E-mail: m.f.craciun@exeter.ac.uk

**The growth of graphene using resistive-heating cold-wall CVD is demonstrated**. This technique is 100 times faster and 99% lower cost than standard CVD. A study of Raman spectroscopy, atomic force microscopy, scanning electron microscopy and electrical magneto-transport measurements shows that cold-wall CVD graphene is of comparable quality to natural graphene. Finally, the first transparent flexible graphene capacitive touch-sensor is demonstrated.

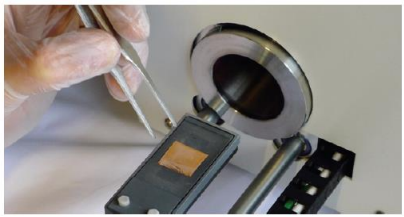 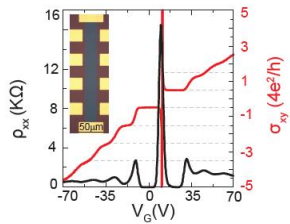 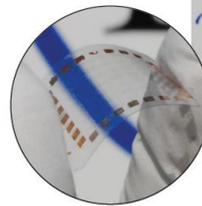 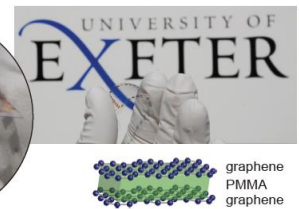

Chemical vapour deposition (CVD) of monolayer graphene on copper[1,2] has emerged as one of the most competitive growth methods for securing the industrial exploitation of graphene, due to its compatibility with Si and roll-to-roll technologies.[3] Recently, there has been tremendous progress in controlling the morphology,[4–6] functionalization[7–10] and growth of heterostructures of intrinsic and doped graphene.[11] However, the low-throughput and the very high production cost for high-quality CVD graphene are central challenges for the industrial exploitation of this material.[12,13] The most common CVD approach is to use a hot-wall system where Cu foils are heated at temperatures ≈1000ºC in a quartz tube furnace through which the precursor hydrocarbon gas flows. The long processing time, that can take a few hours, limits the throughput of graphene by this method. At the same time the typical cost of graphene produced in this way is in excess of £1/cm$^2$, whereas its retail price ranges from £4.57/cm$^2$ to £21/cm$^2$ (see Supporting Information). Therefore, a way forward to increase the throughput and reduce the production cost is to grow graphene in a cold wall CVD system which heats selectively only the Cu foils. Few types of cold wall CVD have been investigated so far for the growth of graphene[3,14–19] such as magnetic induction heating CVD,[14] rapid thermal annealing CVD using halogen lamp heating,[15,16] Joule heating CVD[17,18] and resistively heated stage CVD.[19] Of all these methods, the resistively heated stage CVD approach allows for faster, more efficient heating and cooling, shorter growth time and less gas consumption. This method provides a more uniform substrate heating, it reduces the chemical reactions which can take place in the gas phase at high temperature known to contaminate graphene and it allows for very fast cooling rates, which have been shown to enhance the quality of graphene grown by CVD on copper foil .[20] Furthermore, this type of cold-wall CVD system is found in manufacturing plants of the semiconductor industries. Most importantly we show that with this method truly high quality monolayer graphene can be reproducibly grown. To date, virtually nothing is known on the growth mechanism of monolayer graphene by cold-wall CVD, as well as on its quality and suitability for flexible electronic applications. Therefore, understanding the growth and properties of graphene obtained with cold-wall CVD is imperative to enable the exploitation of this material and facilitate the birth of novel graphene-based applications.

Here we report a completely new mechanism for the growth of graphene by resistively heated stage cold-wall CVD which is markedly different from the growth mechanism of graphene in a hot-wall CVD. Through a combined study of Raman spectroscopy, atomic force microscopy (AFM) and scanning electron microscopy (SEM) we elucidate the early stage formation of graphene by monitoring the transition from disordered carbon adsorbed on Cu to graphene. We also demonstrate for the first time (1) high-throughput production, (2) ultra low cost and (3) high quality monolayer graphene grown on Cu foils by resistively heated stage cold-wall CVD. Our technique merges short deposition time (≈ few minutes) with high-efficiency heating of a cold-wall CVD system, resulting in ≈ 99% reduction in graphene production cost. The Raman spectra of our graphene films shows a low defect related peak and in devices with an area of 5600 µm$^2$ fabricated on standard SiO$_2$ substrates we measure a charge carrier mobility of 3300 cm$^2$V$^{−1}$s$^{−1}$ and the quantum Hall Effect typical of single layer graphene. In contrast, the quality of graphene grown by hot-wall CVD is often gauged only by carrier mobility,[1,2,4,5,21,22,23] giving little information regarding the large area properties of the film. Therefore, to better quantify the quality of graphene films for electronic applications, we introduce an electronic quality factor (Q) accounting for the area across which the carrier mobility is measured. Using Q as a gauge we show that graphene grown by cold-wall CVD has enhanced quality compared to the material grown by hot-wall CVD. Finally, we demonstrate that graphene grown by cold-wall CVD is suitable for the next generation electronics by embedding it into the first transparent and flexible graphene capacitive touch-sensor that could enable the development of artificial skin for robots.

Studies of the growth mechanism of graphene on copper (using methane) in a hot-wall CVD have thus far suggested the direct growth of two-dimensional films involving several steps. The first step is the direct formation of two-dimensional nuclei of graphene[24] from the adsorbed carbon species resulted from the catalytic decomposition of methane on the copper surface. These graphene nucleation sites subsequently grow with the addition of carbon to their edges to form islands and large domains.[25] The growth parameters such as the temperature, pressure, growth time and gas flow are tuned to let graphene domains grow until they coalesce and a continuous graphene film is attained.[26] Though it has been suggested that after the growth of the first layer the catalytic copper surface becomes passivated and limits the growth of other layers, several studies of low-pressure CVD have reported the growth of bilayer[27] and trilayer,[28] as well as multilayers for atmospheric pressure CVD.[29] Nevertheless, the thickness of the grown layers in a hot-wall CVD is always limited to few nanometers or less.

Our experiments show that the growth mechanism of graphene in cold-wall CVD is markedly different from that of the hot-wall CVD described above. Specifically, over a range of growth temperatures that we have investigated, we always observed a thick carbon film (100nm), which forms in the early stages of the growth (see Figure 1a, top left), that becomes progressively thinner with increasing the growth time (see top inset in Figure 1b) and finally evolves into graphene islands (see Figure 1a, top right). The time required to form graphene decreases from 6 minutes at 950°C to 20 seconds at 1035°C (see Supporting Information). To elucidate the initial stage of graphene growth, that is the adsorption of carbon on the Cu substrate, we focus on the slow graphene formation at 950°C. Graphene films were obtained using a commercial cold-wall CVD system (see Supporting Information for details on the design and stability of critical parameters needed for the growth of high quality graphene with this process). The films were transferred from the Cu foils to $SiO_2$/Si substrates using a wet transfer method.[1,30] Full details of the growth and transfer procedures are provided in the Supporting Information. Similar studies for films grown at higher temperatures are presented in the Supporting Information.

Figure 1b shows the Raman spectra of films grown at 950°C for growth time ($t_G$) ranging from 1 to 6 minutes. For all the samples we observe the characteristic peaks of sp2 bonded carbon atoms: the D-peak at ≈ 1340 cm−1, the G-peak around 1600 cm−1, the D'-peak around 1620 cm−1 and the 2D-peak at ≈ 2700 cm−1. For short $t_G$ (i.e. 1 to 4 minutes) the D- and G- peaks have considerable higher intensities than the 2D-peak, which is typical of disordered carbon films.[31,32] As $t_G$ increases we observe changes in intensities, sharpness and positions of the D and G peaks, and for $t_G$>4 minutes a well-defined 2D-band emerges. At the same time, AFM measurements show a reduction in the film thickness from 116 nm to 2.7 nm with increasing $t_G$ from 1 to 6 minutes (see top inset of Figure 1b), which suggests the desorption of carbon from the film.

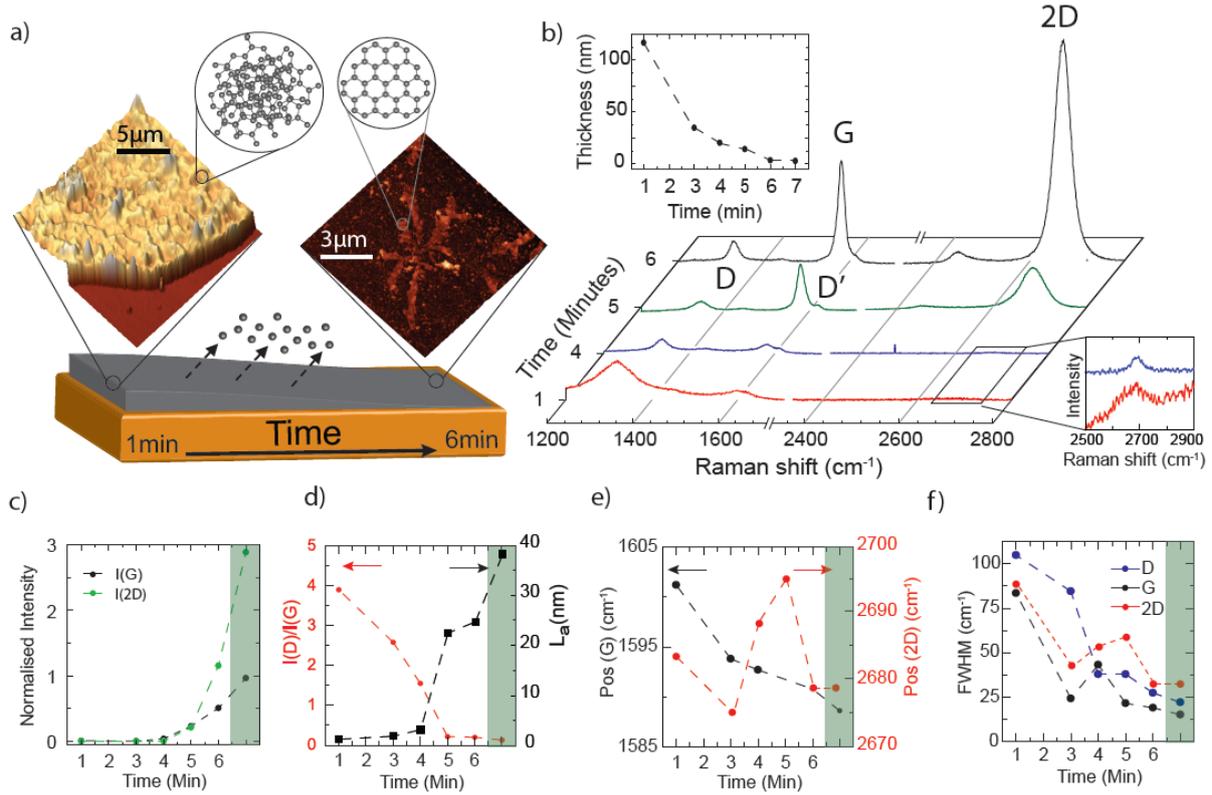

**Figure 1.** a) An illustration of graphene formation from a disordered carbon film. The inset images are AFM topographies for a disordered carbon film (left) and a graphene island (right). The top insets show schematic crystal structures of disordered carbon (left) and graphene (right). b) Raman spectra of films grown at 950◦C for different growth times and transferred to SiO2/Si. The peaks are normalised to the Si-peak intensity. The top inset shows the film thickness as a function of growth time measured using AFM. The bottom inset shows the presence of 2D bands for $t_G$<4 minutes. c) The intensity of the G- and 2D-peaks normalised to the Si-peak intensity as a function of growth time. d) The ratio of D-peak intensity to G-peak intensity and the sp2 cluster size (La) plotted as a function of growth time. e) shows the evolution in G- and 2D- peaks position as a function of growth time. f) shows the reduction in the FWHM of the D- and G- and 2D peaks as a function of growth time. The green regions in c) to f) indicate values taken from a continuous monolayer graphene film.

Lorentzian fitting of the D-, G- and 2D-peaks allows us to ascertain the structural ordering within the films by analyzing the band intensities ($I_{D,G,2D}$), the Full Width at Half Maximum (FWHM(D,G,2D)) and the peak positions (Pos(G,2D)). According to the three stage model for classification of disorder,[33–37] the evolution of $I_D/I_G$, FWHM(D,G) and Pos(G) allow us to assess the ordering/amorphization in carbon materials ranging from graphite and amorphous carbon[33,34] to few-layer and monolayer graphene.[35–37] For $t_G$=1 min, the presence of a 2D peak with Pos(2D)=2683 cm$^{-1}$ and FWHM(2D)=88 cm$^{-1}$, the absence of a doublet in the D and 2D peaks, together with the overlap of G and D' peaks indicate the formation of nanocrystalline graphite with no three-dimensional ordering. Figure 1c shows that $I_{2D}$ and $I_G$ increase with increasing $t_G$, whereas the ratio $I_D/I_G$ decreases from ≈ 3.9 to 0.2 (see Figure 1d). At the same time Pos(G) down-shifts from 1601 cm$^{-1}$ to 1590 cm$^{-1}$ (see Figure 1e) and a significant reduction of FWHM(D,G) occurs (see Figure 1f). The evolution of $I_{G,2D}$, $I_D/I_G$, Pos(G) and FWHM(D,G) with increasing $t_G$ is consistent with the stage 1 ordering trajectory leading from nanocrystalline graphite to graphite. In this regime the size of

sp$^2$ clusters (L$_a$) increases with increasing ordering and can be estimated using the Tuinstra-Koenig relation I$_D$/I$_G$=C($\lambda$)/L$_a$ where C(532 nm)≈4.96nm.[38,39] Using this relation we estimate L$_a$≈2nm for t$_G$=1 min, which increases to L$_a$≈25nm for t$_G$=6 min as shown in figure 1d.

For t$_G$ >6 minutes the 2D-peak intensity is larger than two times the intensity of the G-peak and it can be fitted with a single Lorentzian, with Pos(2D)=2678 cm$^{-1}$ and FWHM(2D)=30 cm$^{-1}$ indicating the formation of monolayer Graphene.[35,40,41] This conclusion is supported by AFM measurements showing the formation of islands with a thickness of 2.7nm, which corresponds to monolayer graphene and accounts for fabrication residues and substrate effects.[42] Furthermore, electrical transport measurements performed on continuous films with a similar Raman spectra and AFM thickness show the quantum Hall effect typical of monolayer graphene (see Figure 3).

To provide further insights into the transition from nanocrystalline graphite film to graphene islands we monitor the evolution of the density, size and separation of the islands using SEM observations combined with a simple counting algorithm described in the Supporting Information. Figure 2a shows the evolution from a continuous film to discrete islands with increasing growth time for 950°C. These images have been performed on the same samples used for the Raman measurements in Figure 1. The average island area within the same range of growth times is shown in Figure 2b, whereas the average separation between islands at initial fragmentation then from 4 to 10 minutes is shown in Figure 2c. An initial reduction in island size suggests desorption of material from the surface. The observed saturation in the island separation of 7.23 μm indicates that there is no further nucleation of islands after the initial fragmentation. After 7 minutes we see a maximum in island size of 19.7 μm$^2$. Raman measurements confirm that these islands are composed of graphene. SEM analysis of films grown for 1000°C and 1035°C reveals a similar behaviour of the saturation in island separation and a maxima in island size (see Supporting Information). We observe that an increase in growth temperature leads to a reduction in the time required to achieve the maximum island size and to form a monolayer graphene as shown in the inset of Figure 2c. A similar behaviour has been also observed in other CVD graphene growth studies,[24,26] which showed that the growth rates of graphene islands are determined by competing atomic phenomena such as adatom mobility and attachment to the islands edges versus desorption, as well as being affected by the microscopic substrate roughness.[26] The counterintuitive decrease in island area with time can be understood within the desorption controlled regime[26] where the growth is a thermally activated process with a barrier energy E$_a$= (E$_{des}$ + E$_{att}$ - E$_d$ - E$_{ad}$)/2 and with the density of graphene islands N$_i$~ P$_{CH_4}$ · exp(2E$_a$/KT), with E$_{des}$ the desorption energy of a carbon monomer on the Cu surface, E$_{att}$ the barrier of attachment for the capture of a monomer by supercritical nucleus, E$_d$ the activation energy of surface diffusion of a monomer, E$_{ad}$ the activation energy for dissociative adsorption of CH$_4$ on Cu, P$_{CH_4}$ the methane partial pressure, K the Boltzmann constant and T the growth temperature. Figure 2d shows that when the island area decreases with time, N$_i$ has a dependence on growth temperature which is typical of the desorption controlled regime with an activation energy of 1.66 eV. The desorption model is also consistent with the formation of holes inside the islands at 8 minutes of growth (see Fig 2a).

The observed transition from a disordered carbon film adsorbed on Cu to graphene is very likely due to the combination of high temperature, low pressure and the presence of the catalytically active surface of Cu, which induces the conversion to graphene as well as the thinning process of the carbon film. Previous studies [43-45] have also investigated the high temperature conversion of amorphous carbon (a-C) films into graphene. In-situ transmission electron microscopy (TEM) and molecular dynamics (MD) studies [43] have reported the high temperature conversion of amorphous carbon (a-C) into graphene patches of 100 x 300 nm$^2$. It was shown that a-C can rearrange into graphene through a phase of glasslike carbon which takes place within a time frame from 1 to 15 minutes, in the temperature range of 326°C – 926ºC. Another study [44] showed that graphene can be grown in a solid-state transformation of a-C in the presence of a catalytically active metal at temperature up to 720ºC. In this case rearrangement processes take place in two or three dimensional unordered network structures in which a huge number of bonds are broken and newly formed. Finally, a third study showed the metal-catalyzed crystallization of a-C to graphene by thermal annealing at 650–950 °C. [45] It was shown that part of the carbon source is crystallized into graphene with the rest outgassing from the system. Furthermore, this study also reports that for long annealing times no carbon or graphene remains on the surface due to significant desorption of C atoms under the low pressure and high temperature ambient. Similarly to these studies we have a film of nanocrystalline graphite on top of a catalytically active metal in low pressure and high temperature conditions, as well as comparable time frames for the conversion to graphene.

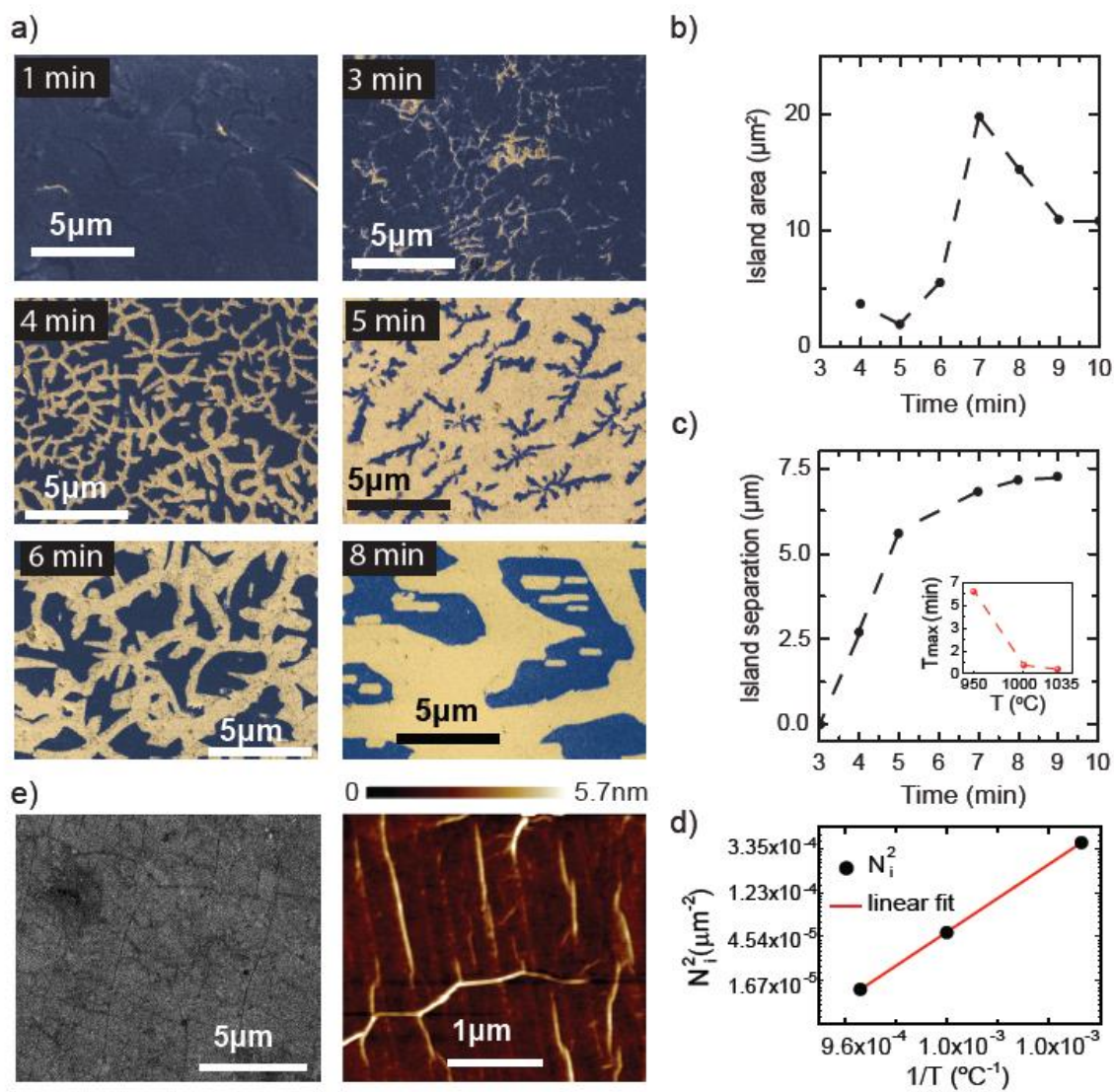

**Figure 2.** a) SEM micrographs sampled over a time frame of the transition from nanocrystalline graphite to graphene. The dark blue color corresponds to graphene, whereas the yellow color is the substrate. b) The evolution of the average island area with the growth time. c) The average separation of islands as a function of growth time. The inset shows the time to reach maximum island size $T_{max}$ plotted as a function of growth temperature. d) Graphene island density as a function of inverse of growth temperature in the regime where the island area decreases with growth time. The red line is a fit to the desorption controlled regime model. e) SEM (left) and AFM (right) images of the continuous monolayer graphene films.

Having established the initial stages of graphene formation, we investigate the transition from graphene islands to a continuous film. Fig. 2c shows that the island size reaches a maximum with the growth time and a further increase in the growth time leads to a decrease in the island size. To grow continuous graphene monolayer films we adopted the two stage growth described by Li et al.,[26] where increasing methane flow rate after the formation of the islands is shown to fill the regions between islands while suppressing further nucleation sites. As our objective is to minimize growth time, we selected the growth temperature of 1000°C where maximum island size and island separation are reached in the shortest time (40 seconds). Using the grown graphene islands as nucleation sites, we find that increasing the methane flow rate and growth time to 5 minutes allows the islands to merge into a continuous graphene monolayer film of up to 8 cm$^2$ in area. SEM, AFM and Raman measurements confirm that the continuous films are monolayer graphene. Figure 2e shows the morphology of the graphene monolayer after the complete coalescence of the islands studied by SEM and AFM. The analysis of the Raman measurements performed on the continuous films is presented in Figure 1, where the green highlighted regions in panels c) to f) indicate the values of $I_{G,2D}$, $I_D/I_G$, $L_a$, Pos(G) and FWHM(D,G,2D) for a 1 × 1 cm graphene film. Raman mapping measurements shown in Supporting Information demonstrate the uniformity and high-quality of the continuous films. The total processing time of this procedure is about 20 minutes (see Supporting Information); this includes (1) heating up time for the CVD system from room temperature to the growth temperature, (2) Cu foil annealing time, (3) graphene nucleation and growth time, (4) cooling down time for the system to room temperature. The demonstrated processing time is significantly shorter than the processing time needed by hot-wall CVD (typically > 70 minutes).[1,2,4,5,21,22] We estimate the total cost of graphene production by cold-wall CVD to be < £0.37/cm$^2$ (see Supporting Information). Compared with other CVD studies and neglecting the base cost of copper we see a reduction in the production costs of 98.83% - 99.89%. This extraordinary reduction in the production cost, together with the possibility of reconstitution of high purity copper from etchant solutions by electrolysis that can yield up to 99% of the original foil,[46] open a new way forward to accelerate the commercialization of graphene.

To ascertain the quality of the electronic properties of graphene produced by cold-wall CVD we characterized the charge carrier mobility in transistor devices fabricated on SiO$_2$/Si substrates. Using the parallel plate capacitor model we estimate the field effect mobility to be 3300 cm$^2$V$^{-1}$s$^{-1}$ at 1.4K and 2773 cm$^2$V$^{-1}$s$^{-1}$ at room temperature. This mobility, measured across a large area device (≈ 0.05 mm$^2$), is comparable to the mobility measured in smaller area devices of graphene either grown by CVD (50μm$^{2[1]}$ to 0.03 mm$^{2[23]}$) or mechanically exfoliated (typically few μm$^2$) and deposited on oxidized silicon substrates.[1,2,4,5,21,22,47–49] The quality of cold-wall CVD graphene as compared to that grown with other methods is readily assessed using the electronic quality factor (Q) that, for the area across which the carrier mobility is measured, is defined as the field effect mobility (cm$^2$V$^{-1}$s$^{-1}$) multiplied by the area of the device (μm$^2$). As shown in the Supporting Information, graphene grown by resistive heating cold-wall CVD has Q ranging from 4×10$^6$ to 7.2 × 10$^6$, whereas most reports of monolayer graphene grown by hot-wall CVD have Q ranging from 10$^3$ to 7 × 10$^6$. Hence cold-wall CVD grown graphene has a narrow spread of Q stemming from a reproducible high quality growth process. This is in stark contrast to the spread of Q over 3 orders of magnitude reported for hot-wall CVD grown graphene.

Figure 3a shows the four-terminal resistance measured in a large Hall bar geometry (225 μm$^2$ x 25 μm$^2$, see inset) fabricated on standard SiO$_2$/p-Si substrates. The Si substrates is heavily doped and acts as the gate electrode. By applying a voltage to the gate (Vg) we tune the carrier concentration from 3x10$^{11}$ cm$^{-2}$ to 6x10$^{12}$ cm$^{-2}$. The charge neutrality point is at 0V, indicating low residual doping levels in our samples. Figure 3b shows the resistivity and the Hall conductance (σ$_{xy}$) against the applied V$_g$, taken at 13T and at a temperature of 250mK.

Clearly developed conductance plateaus are visible when the Fermi energy is within a Landau level (N) that correspond to the conductance relationship $\sigma_{xy} = (N + 1/2)4e^2/h$, typical of the half-integer quantum Hall effect (QHE) of monolayer graphene with 2-fold spin and valley degeneracy.[48,49] At the same time we observe $\rho_{xx}= 0$ Ω where the Fermi energy is within the N=0 Landau level for both electrons and holes, indicating that the graphene quality is high enough to observe the localisation in the QH regime. At the same time, $\sigma_{xx}$ shows well-defined Shubnikov-de Haas oscillations which are periodic with the applied gate voltage. A color map of the differential Hall conductance plotted against perpendicular applied field and charge carrier concentration shows that Landau levels up to N=6 are visible at high fields (see Figure 3c). The N=1 Landau level is visible down to fields as low as 5T. The presence of these QHE features at low fields is an indication of low disorder in the graphene which further demonstrates the high electronic quality of cold-wall CVD graphene.

In the final section of this article we demonstrate that graphene produced by this novel method is suitable for the next generation flexible and transparent electronics. In such applications touch sensing is the dominant human interface method for detecting an input. Among the touch sensing devices, the capacitive touch sensors have the fastest response time and the best sensitivity to touch input. However, graphene-based flexible capacitive touch sensors have not been demonstrated so far due to the difficulties arising from poor adhesion of subsequent graphene and dielectric layers on flexible substrates. We developed a novel fabrication procedure that preserves the high quality of graphene (see Supporting Information), therefore allowing us to demonstrate for the first time a flexible and transparent capacitive touch sensor using graphene for both the top and bottom electrodes.

Figure 3d shows a photograph of a capacitive touch sensor array fabricated on a flexible and transparent Polyethylene naphthalate (PEN) substrate. The array consists of two orthogonal sets of graphene strips separated by PMMA dielectric as illustrated in Figure 3d. The individual graphene strips were fabricated on the Cu foil and connected to 50 nm thick Au pads, followed by their transfer to the PEN substrate. Electrical transport measurements show that the typical resistivity across each strip length is $\rho \approx 1.3$ kΩ and the contact resistance < 68 Ω. A detailed description of the fabrication procedure and electrical characterization is provided in the Supporting Information. The individual elements of the touch sensors are formed at the intersection between the graphene strips and each element represents a parallel plate capacitor. As pressure is applied to an element of the array, the dielectric elastically deforms reducing the spacing between the graphene electrodes resulting in an increase in capacitance. Figure 3e shows an interpolated colour map of percentage change in capacitance for each element when one element is loaded with a 36 g mass. The maximum change in capacitance occurs on the loaded element with minimal changes to the surrounding elements.

To test the responsivity of the device we periodically loaded and unloaded an element with a human finger and measured the change in capacitance shown in Figure 3f. A change in capacitance during loading of $\Delta C = 6$ pF was observed with a return to the original state after unloading. The sharp change in capacitance demonstrated indicates a fast responsivity to loading and unloading of the element. Finally we tested the flexibility and durability of the devices by bending the substrate and systematically measuring the resistance of the graphene strips across the device. Figure 3g shows the percentage change in the two terminal resistance of the graphene strips as the device is flexed though a 2.5cm bending radius for 2000 iterations. This test was performed for graphene strips parallel (black) and perpendicular (red) to the axis of device flexing. After 2000 bends only minor changes of less than < 2.7% in the line resistance are observed, which show no significant deterioration of the operation of the flexible touch sensor. These measurements demonstrate the sensitivity and durability of our graphene touch sensor, and its suitability for use in next-generation flexible portable devices.

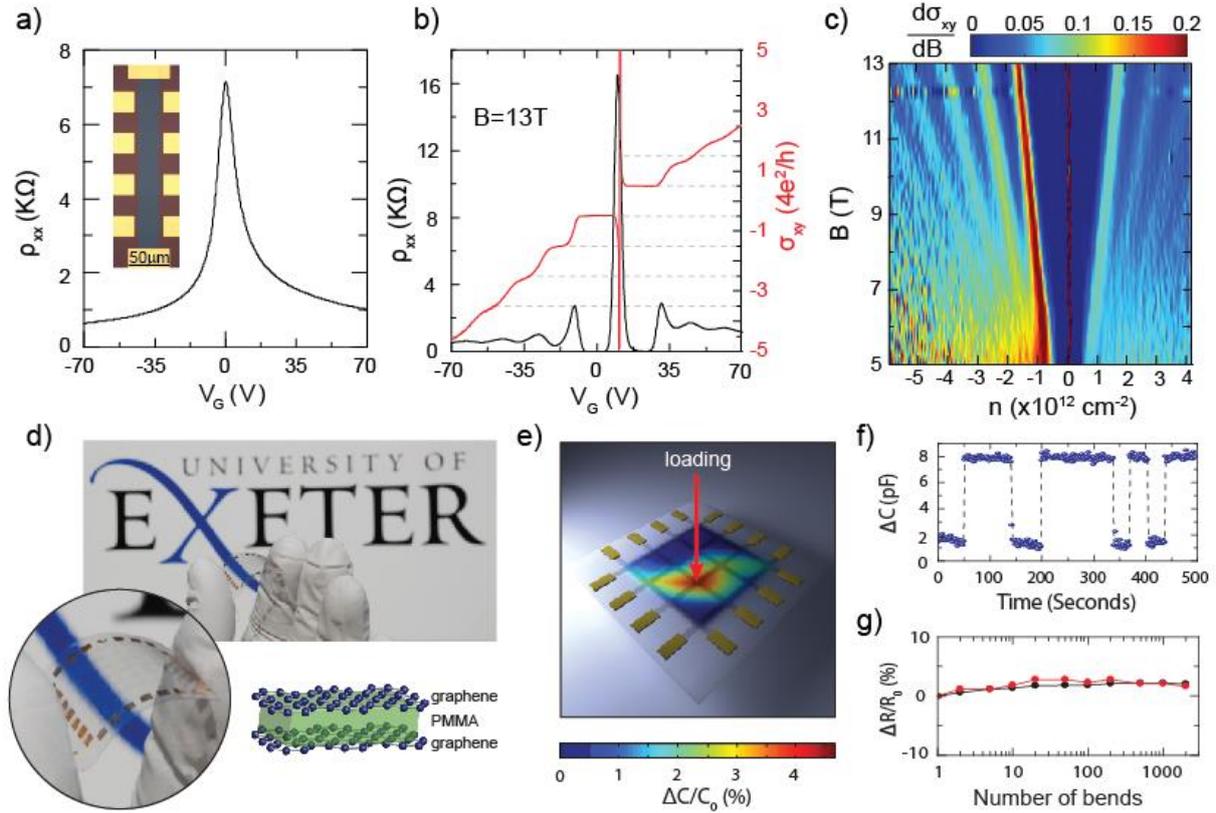

**Figure 3.** a) The longitudinal resistivity ($\rho_{xx}$) plotted against applied gate voltage at 4.2K. The inset shows a false colour photograph of the device. b) The longitudinal resistivity and the Hall conductance ($\sigma_{xy}$ normalised to $4e^2/h$) plotted against applied gate voltage at 250mK with a 13T perpendicular applied field. c) Colour map of the differential conductance as a function of applied perpendicular magnetic field and carrier density. d) Shows a photograph of a flexible and transparent graphene touch sensor and a schematic of the touch sensor device. e) Colour map of the change in capacitance when a single element is loaded with a 36g mass. f) The change in capacitance of one element with respect to time when pressed with a human finger. g) The change in line resistance after flexing the device about a 2.5 cm radius. Black and red points show the resistance of line parallel and perpendicular to the bending radius respectively

In summary, we have shown a new growth mechanism of graphene by cold-wall CVD, which starts with the formation of a thick carbon film in the early stages of the growth, that becomes progressively thinner with increasing the growth time and finally evolves into graphene islands. At the same time we demonstrate an extremely high-throughput and cost efficient growth procedure for preparing high quality monolayer graphene using cold-wall CVD. Finally, we use graphene as electrode material and demonstrate the first flexible and transparent graphene capacitive touch sensor using processing techniques that are compatible with existing transparent and flexible electronic technologies. Besides its importance for the quick industrial exploitation of graphene since cold-wall CVD systems are found in semiconductor industries manufacturing plants, our work could lead to new generations of flexible electronics and offers exciting new opportunities for the realization of graphene-based disruptive technologies.

**Acknowledgements**
We acknowledge financial support from EPSRC (Grant EP/J000396/1, EP/K017160,

EP/K010050/1, EP/G036101/1, EP/M002438/1, EP/M001024/1) and from the Royal Society Travel Exchange Grants 2012 and 2013.

Supporting Information

**High quality monolayer graphene synthesized by resistive heating cold wall chemical vapour deposition**

*Thomas H. Bointon, Matthew D. Barnes, Saverio Russo, and Monica F. Craciun\**

*Description of the cold-wall CVD system used for the growth of graphene*

The growth of graphene was performed in a commercial cold-wall CVD system from Moorfield (i.e. nanoCVD-8G system). In the following we provide a description of the system as well as the required parameters and schematic representations needed to grow graphene with any resistive heating cold-wall CVD process.

This system is enclosed in a metallic case and consists of a stainless steel vacuum chamber, mass flow controllers for the gas delivery, valves for chamber pressure control and purge, and a gauge for pressure measurement (see Figure S1). The system is equipped with 3 types of process gasses ($CH_4$, $H_2$ and Ar). However in this study we only used $CH_4$ and $H_2$ for the growth. The purge gas is argon.

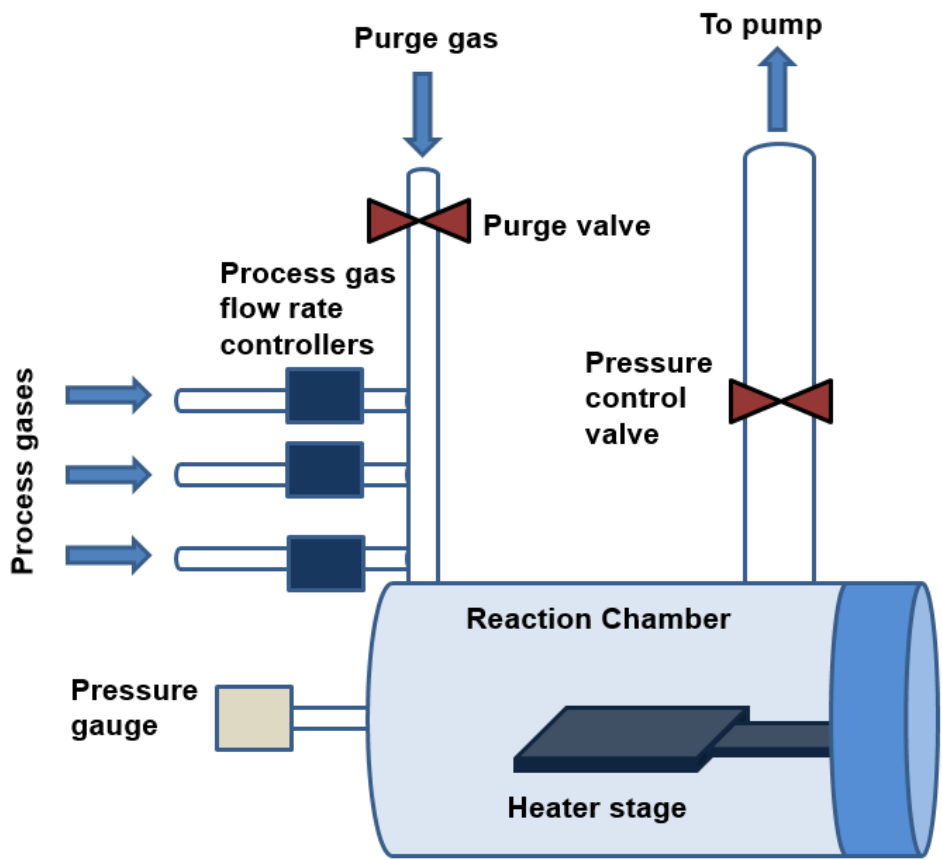

**Figure S1:** Schematic diagram of the cold-wall CVD system used for graphene growth. The arrows indicate the direction of gas flow.

The reaction chamber houses a resistively heated substrate stage equipped with an embedded thermocouple which can achieve stable temperatures of up to 1100ºC. The heater assembly slides out of the chamber for substrate loading (see Figure S2) and is then pushed back in the chamber. The hardware is controlled by a programmable logic controller electronics coupled to a touchscreen interface and all operation of the system is carried out through the touch screen. In this system the Cu foil is placed on the resistively heated stage as shown in Figure S3.

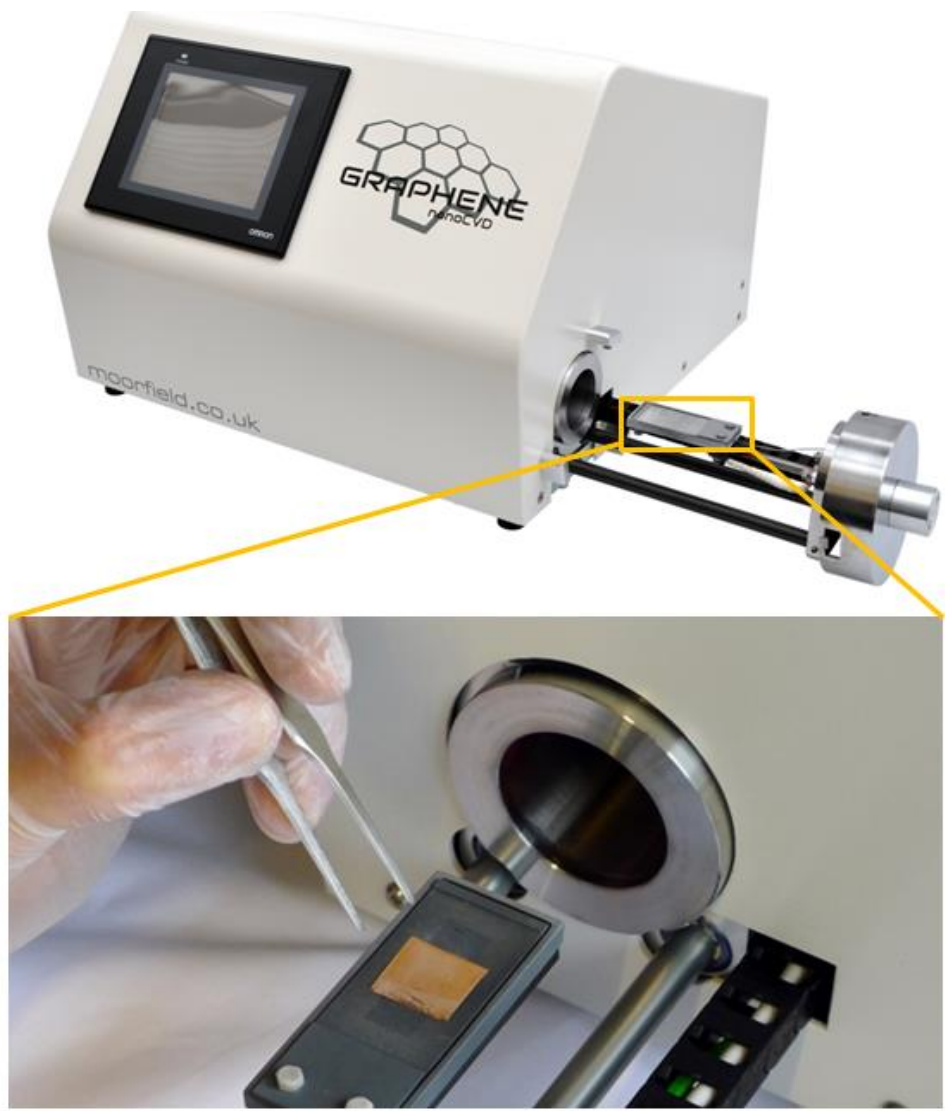

**Figure S2:** A photograph of the CVD system used in our work. The inset shows the resistive element heating stage loaded with a Cu foil [reproduced with permission from the website of Moorfield nanoCVD-8G].

The temperature at the surface of the Cu foil is measured by using a thermocouple mounted on the heater stage, thus in direct contact with the substrate. Figure S11a shows the heater stability for different temperatures as well as the chamber temperature which remains around 100ºC during the heater operation. As the Cu foil is in direct contact with the heater/thermocouple, the temperature of the substrate can be reliably controlled as the introduction of gas does not modify the foils surface temperature (see Figure S3b).

a)

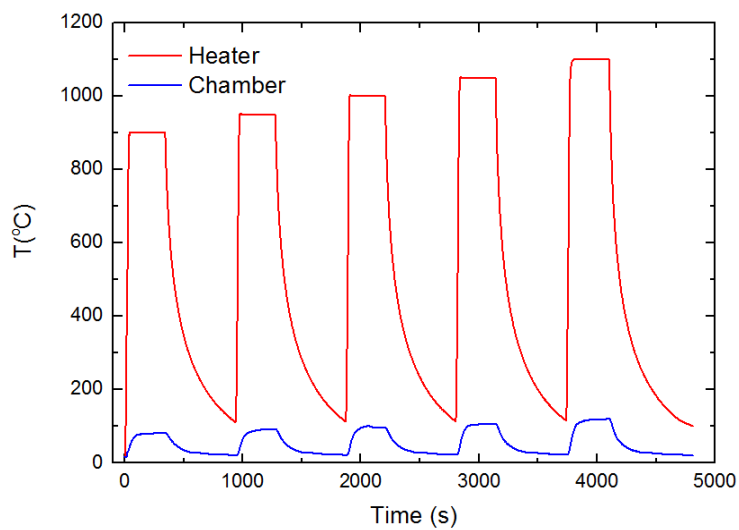

b)

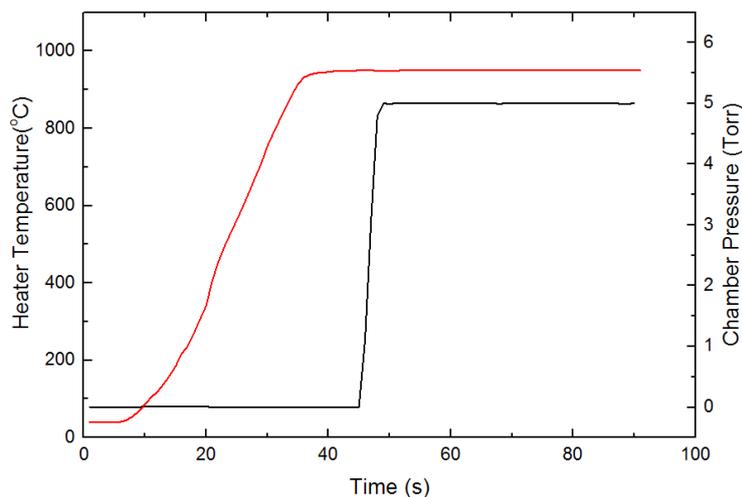

**Figure S3:** a) The stability of the heater temperature (red) in vacuum (P=0.05 Torr) for different temperature set-points ranging from 900°C to 1100°C. The blue curves show the corresponding chamber temperature which is around 100°C. b) The stability of the heater when gas with a pressure of 5Torr is introduced in the system.

The pressure inside the reaction chamber can be reliably controlled using the pressure control valve. Figure S4 shows the pressure stability for different set-points which are achieved in this case by controlling the flow of Ar gas.

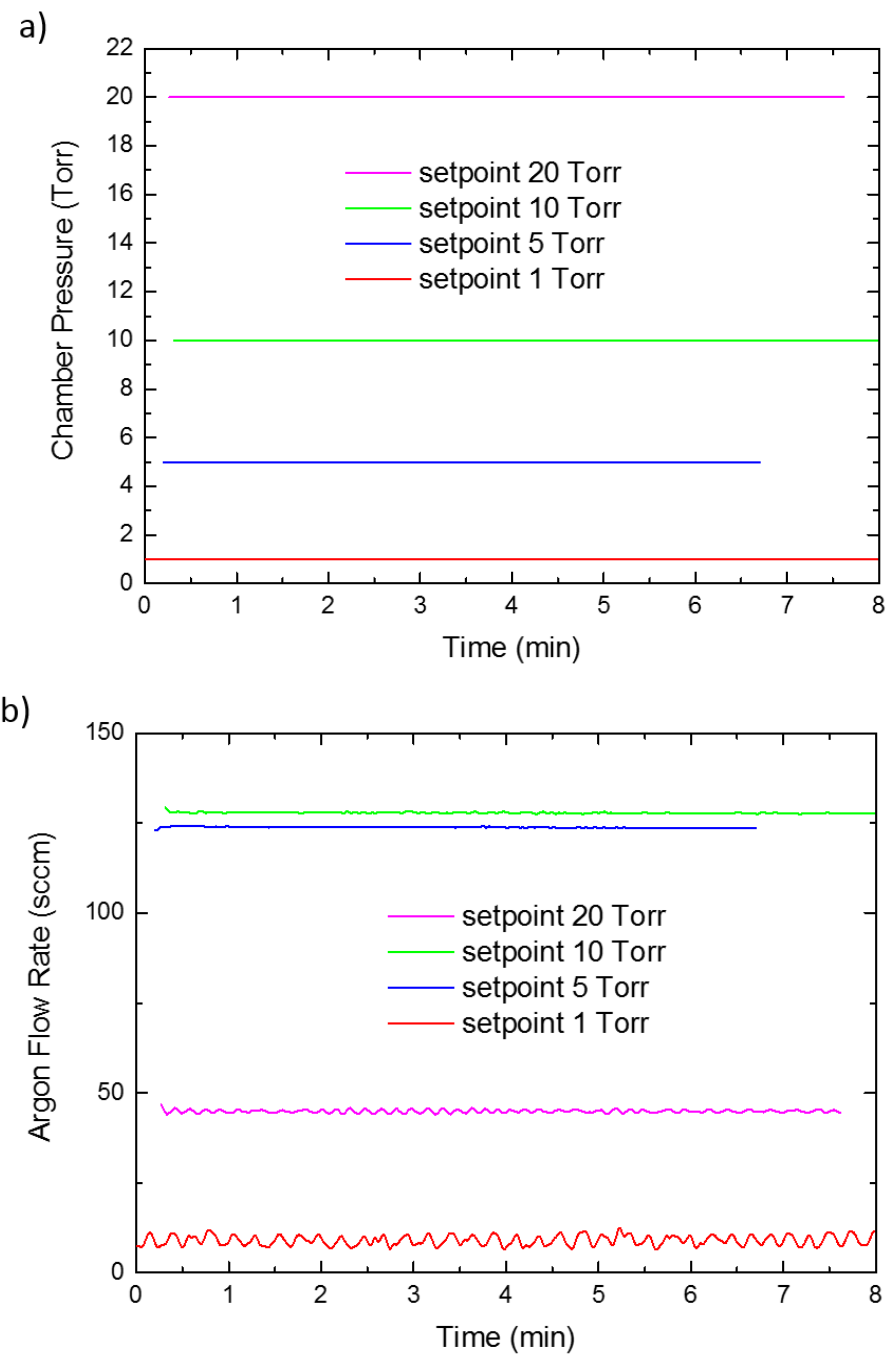

**Figure S4:** Pressure stability for different set-points (a) and the gas flow required to achieve the desired pressure (b).

*Growth procedure for the graphene films and islands*

25 μm thick copper foils (Alfa Aesar 99.999%) were annealed for 10 minutes at 1035°C in a H$_2$ atmosphere to increase the Cu grain size.

To understand the initial stages of graphene formation, the growth was carried out at temperatures ranging from 950°C to 1035°C and the growth time was varied from 10 seconds to 600 seconds. A constant flow rate of 0.4sccm of H$_2$ and 1.4sccm of CH$_4$ was used for all growths.

A typical processing for the growth of continuous graphene films involves the following steps: (1) heating up the CVD system from room temperature to the growth temperature, (2) Cu foil annealing, (3) graphene nucleation and growth, (4) cooling down the system to room temperature (see Figure S5).

During the heating up stage H$_2$ gas was flown at a rate of 0.4sccm with a chamber pressure of 0.01 Torr. The annealing step was performed for 10 minutes at 1035°C in a H$_2$ atmosphere, keeping the H$_2$ gas flow rate at 0.4sccm and the chamber pressure of 0.01 Torr. The temperature was then lowered at 1000°C for the growth of continuous graphene films. A constant flow rate of 0.4sccm of H$_2$ was kept throughout the nucleation and growth. For the nucleation stage, 1.4sccm of CH$_4$ was introduced for 40 seconds. This was followed by the growth stage where the CH$_4$ flow rate was increased to 7sccm for a 300 seconds. Finally, the system was cooled down at room temperature keeping the H$_2$ gas flow rate at 0.4sccm.

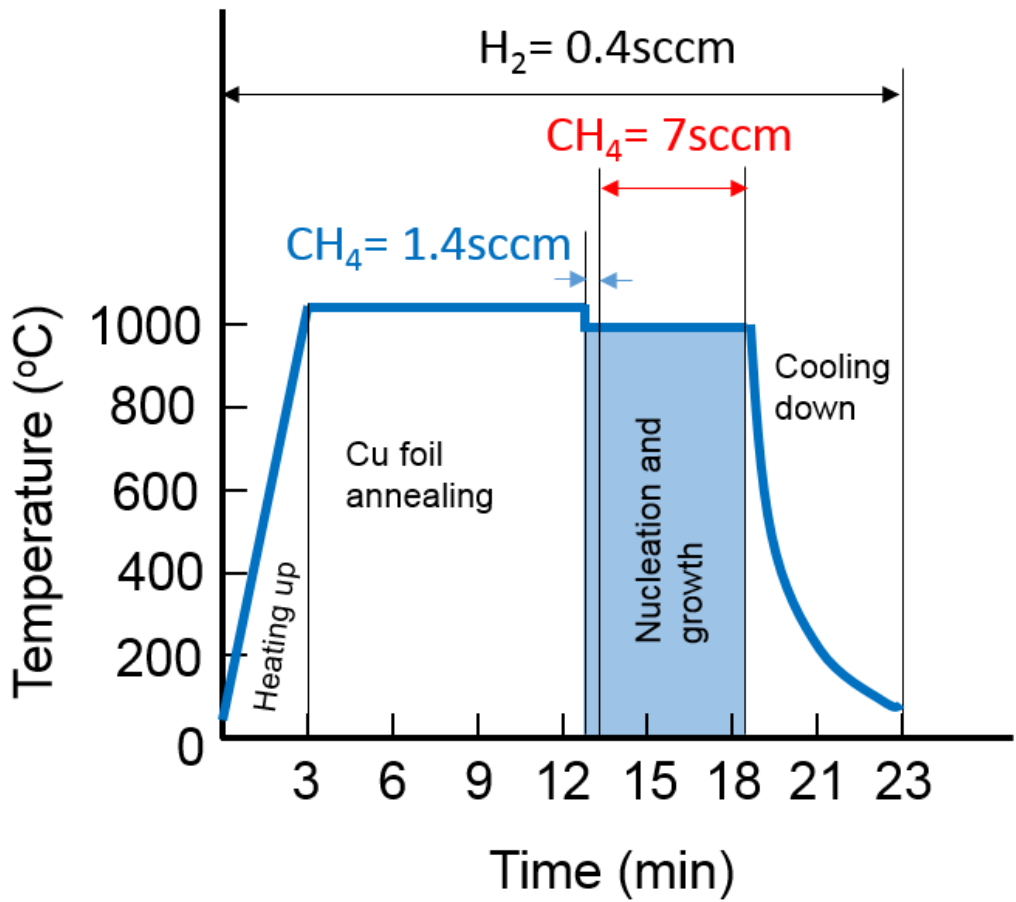

**Fig. S5.** Time dependence of growth parameters for the cold-wall CVD system used in this study.

*Transfer Procedure of graphene films from the Cu foils onto SiO$_2$/Si*

Grown graphene samples were spun with 200nm of 950K PMMA. The PMMA coated foils were vacuum cured for 30 minutes and then etched in 1M FeCl3 solution. After the copper was fully etched the films were transferred several times to deionized water and then transferred onto SiO2/Si substrates.

*Device fabrication*

Graphene devices were produced using standard electron beam lithography and reactive ion etching techniques to define Hall bar geometry (225 μm × 25 μm) shown in the false colour inset of Figure 3a with electrical contacts of Au/Cr (50 nm/ 5 nm).

*Electrical transport measurements*

The longitudinal and Hall voltages were measured in a four terminal geometry applying an AC current using a lock-in amplifier. The excitation voltage was selected to be within the linear transport regime.

*SEM analysis*

SEM Measurements: SEM micro-graphs were collected with a Phillips SEM. An acceleration voltage of 30kV, magnification of x5000 and beam current of 0.63nA was used.

SEM micrographs where taken for graphene islands transferred to $SiO_2$ to determine the average area and separation of domains. Figure S6a shows a micrograph taken at 5000x magnification where graphene islands appear dark and the $SiO_2$ substrate is lighter. The image was then processed by inverting the colors and applying a threshold to create a two colour bitmap shown in Figure S6b. Using the matlab image processing toolbox, each island was identified and the area was measured [1]. Figure S6c demonstrates a single identified island on a false colour map. To reduce the effects of residues resulting from the transfer process the results were filtered to remove any island with an area smaller than 1 µm². The resulting islands were given random false colour to check that no islands are connected as demonstrated in Figure S6d.

All calculations were based on 10 micrographs for each growth time, where the average area was estimated by summing the area of all islands ($A_{islands}$) and dividing by the total number of islands ($N_{islands}$). The average separation was estimated from the density of islands ($S_{mean}$)

$$S_{mean} = \sqrt{\frac{1}{d}} = \sqrt{\frac{A_{total}}{N_{islands}}}$$

where density (d) was taken as the total number of islands ($N_{islands}$) divided by the total area of the micrographs ($A_{total}$).

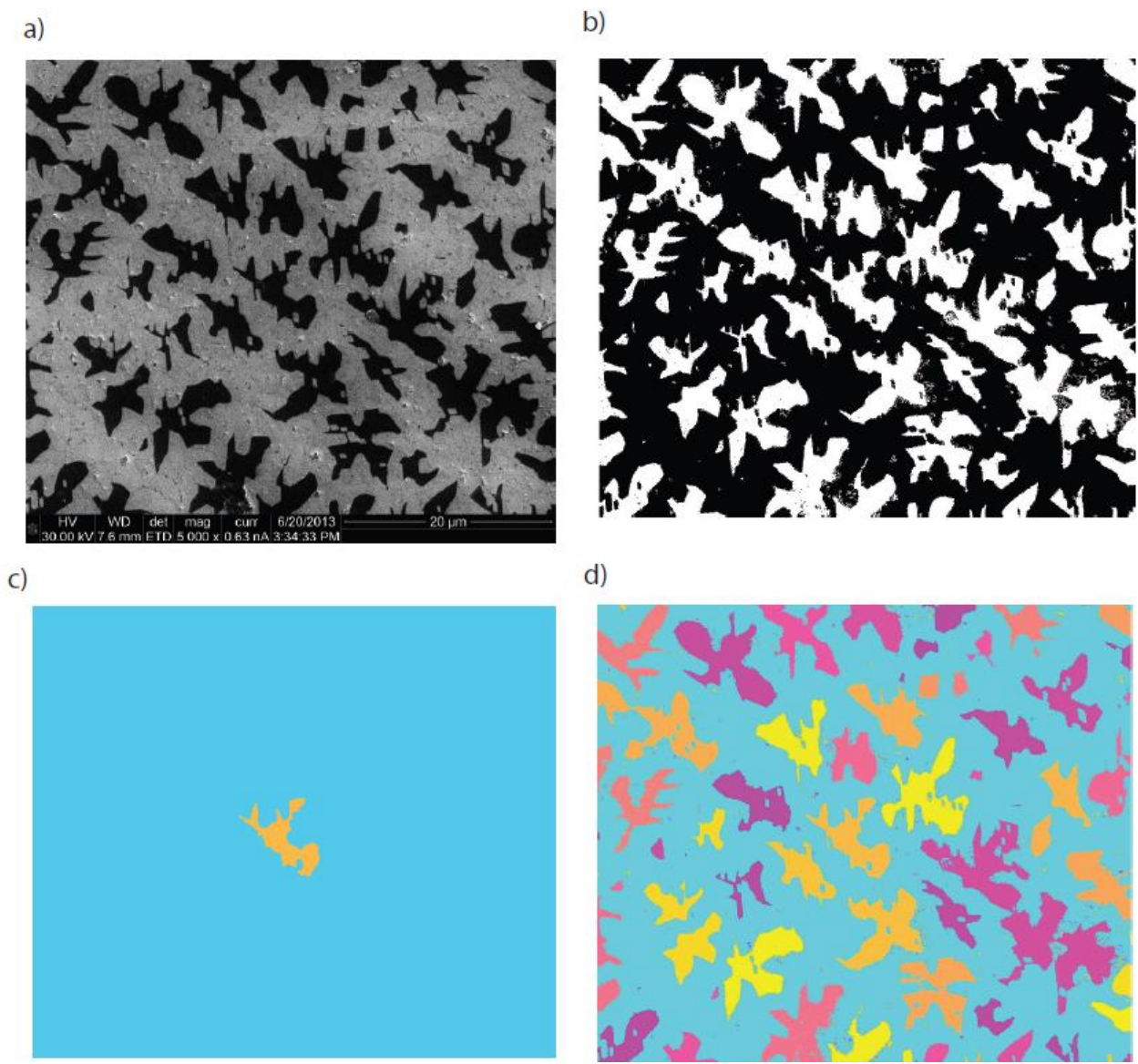

**Figure S6.** a) An SEM micrograph showing Graphene islands (Black) on an $SiO_2$ substrate, b) Processed SEM micrograph with inverted intensities and applied black and white threshold, c) A single identified island extracted from SEM micrograph shown in false colour, d) All identified islands from SEM micrograph after applying noise filter

*AFM analysis*

To study the evolution from a carbon film to graphene islands, semi-contact AFM topography images were collected with a NTMDT Ntegra AFM. Film thickness was extracted by fitting the statistical distribution of the film and substrate heights. For the contious graphene films, the images were colected in contact mode with a Bruker Innova AFM.

The thickness of each growth time was determined using tapping mode AFM where a surface topography was measured, shown in Fig. S7a. An area which includes the substrate and the film/islands highlighted in Fig. S7a was sampled and the distribution was then plotted as a histogram shown in Fig. S7b. The two peaks represent the substrate height and the film/island height. Fitting each peak with a Gaussian and subtracting the height of the substrate from that of the film/islands gives the total film thickness.

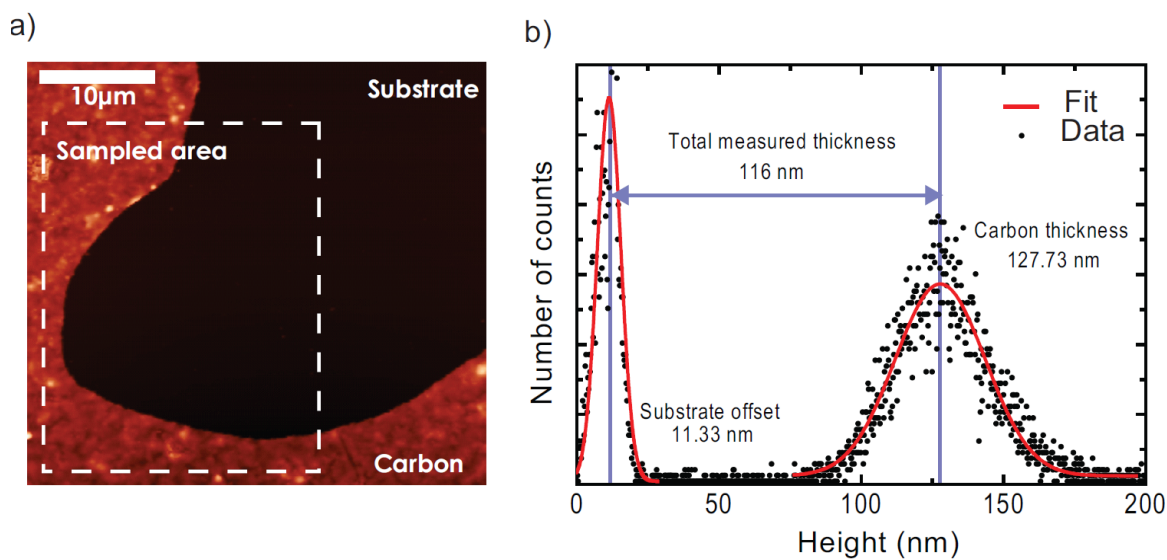

**Figure S7.** a) AFM topography of graphene film on a $SiO_2$ substrate, the highlighted region shows the sampled region for the statistical study. b) A histogram showing the distribution of measured heights from within the sampled area. Substrate and film distributions are fitted and the difference between the average heights gives the total thickness.

*Raman spectra for films grown at 1000°C and 1035°C*

Raman spectra were collected in a Renishaw spectrometer with an excitation laser wavelength of 532 nm, focused to a spot size of 5 μm diameter and x50 objective lens.

For films grown at higher temperatures (1000°C, 1035°C) we observe the same transition from nanocrystalline graphite to graphene islands as for growths at 950°C shown in Figure 1, but at a faster rate. Figure S8a shows several spectra at 1000°C for different times. After 5 seconds of growth we observe the presence of D- and G-bands alongside a small 2D-band characteristic of nanocrystalline graphite. As growth time is increased there is an increase in the intensity of the 2D-band and the ratio of the band intensities of D- and G-band, $I_D/I_G$ transitions from ~2 to <1 demonstrating the increase in long range hexagonal ordering. Furthermore after 40 seconds $I_G/I_{2D} \approx 0.5$ indicating that the islands are monolayer graphene. Similarly Fig. S8b shows several spectra for growth at 1035°C for different times. We observe the spectra of nanocrystalline graphite after 1 second and a transition into monolayer graphene after 20 seconds.

There is a clear relationship between the growth temperature and the time for the nanocrystalline graphite to transition into monolayer graphene, where increasing the temperature reduces the time required to form graphene.

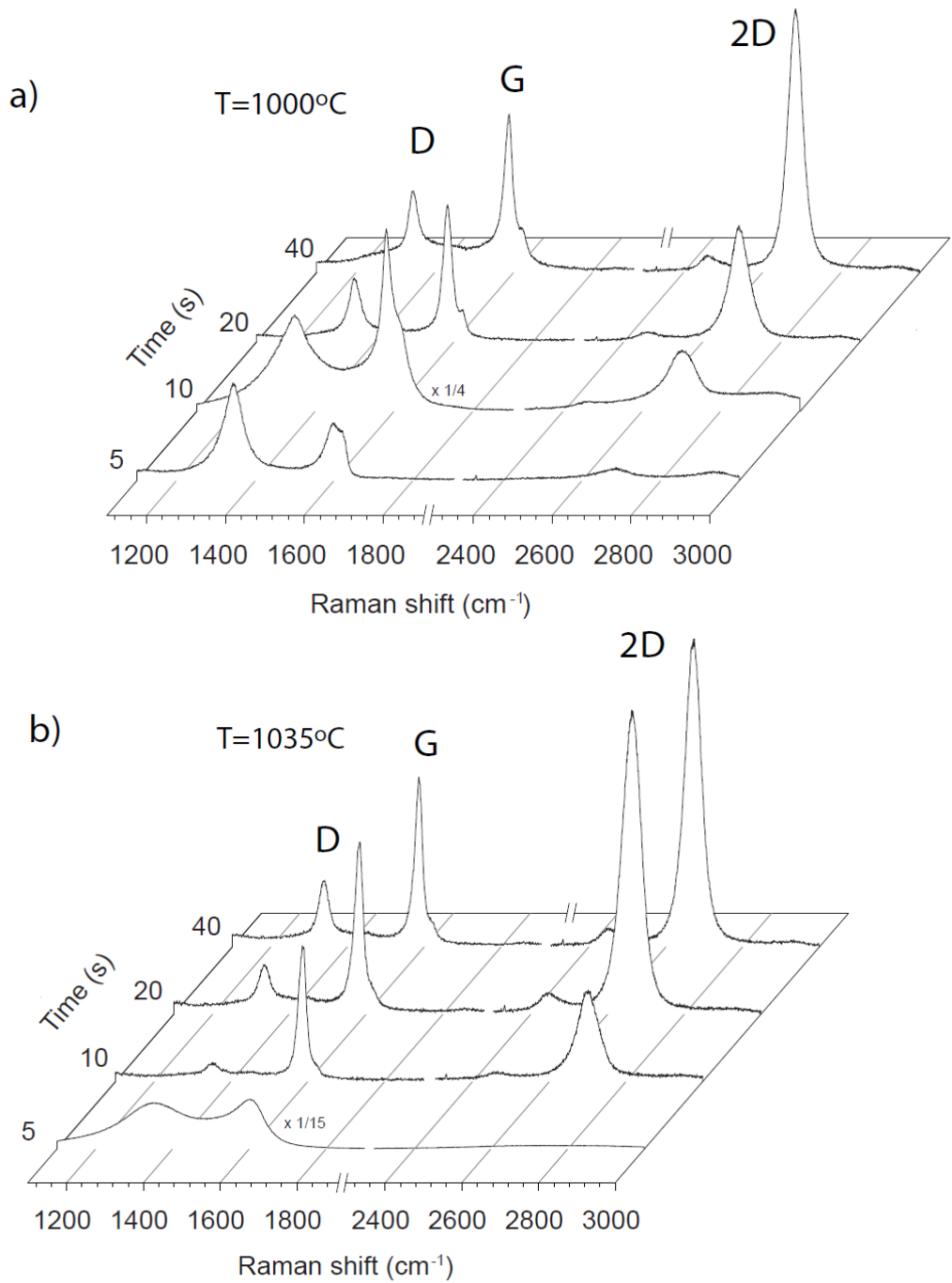

**Figure S8** The evolution with respect to growth time of the Raman spectra for growths at a) 1000ºC and b) 1035ºC.

*Island area and island separation for films grown at 1000ºC and 1035ºC*

Fig. S9a shows the average island area for both 1000ºC and 1035ºC with increasing growth time. The maxima for each temperature corresponds to the same time where we observe a graphene Raman spectra. Increasing the growth time beyond this shows a reduction in the average island area, which correlates to the Raman spectra no longer showing the presence of nanocrystalline graphite. The reduction in area could be due to one of many factors such as hydrogen etching of the graphene [2], the evaporation of the copper substrate under the graphene or a change in the concentration of surface carbon available for growth due to the exhaustion of the carbon [3].

Simultaneously we observe an increase in the average island separation for both temperatures, shown in Fig. S9b. The rate of separation after prolonged growth shows signs of slowing, but it is unclear if the separation will saturate like for the 950ºC growths.

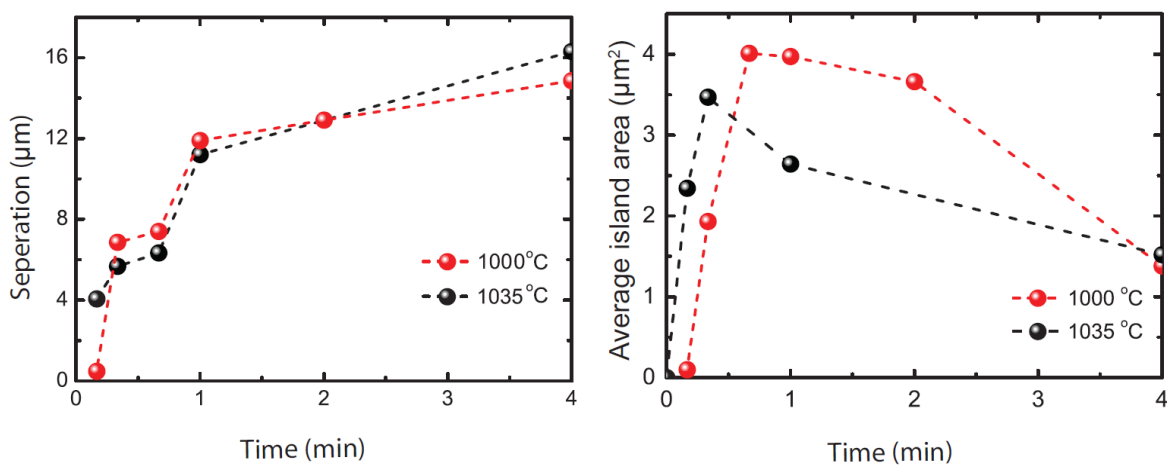

**Figure S9** a) Shows the estimated island separation with increasing growth time for 1000ºC and 1035ºC, b) Shows the estimated average island size with increasing growth time for 1000ºC and 1035ºC.

*Characterization of continuous graphene films*

Fig. S10 shows an optical microscope image (Fig. S10a) and the Raman spectra (Fig. S10b) for three regions of continuous graphene grown using the two stage growth method and transferred on $SiO_2$/Si. The D-, G- and 2D- bands were fitted and used for the continuous growth data points which appear in Figure 1 in the main text. These continuous films were used to fabricate the Hall bar devices shown in Fig S10c (top). Fig. S10c shows the mapping of the Full width at half maximum (FWHM) of the 2D band (middle) and the intensity ratio of the D to G peak, $I_D/I_G$ (bottom). The FWHM of the 2D band ranges from 30 to 35 $cm^{-1}$ which is typical for CVD grown monolayer graphene. The Raman maps have been taken with 1μm step size.

On the continuous films we still observe a small D peak, which indicates defects. However, this peak is usually observed on CVD grown polycrystalline graphene films and it is believed that defects arise from the misalignment of the islands as they come together and coalesce into a continuous film. Indeed, when we grow graphene islands which are larger than the area probed by our Raman measurement (i.e. spot size of 5 μm diameter) we do not observe the presence of the D band as shown in figure S10d. Therefore the D band that appears in the Raman spectra of the films is due to the defects arising from the grain boundaries.

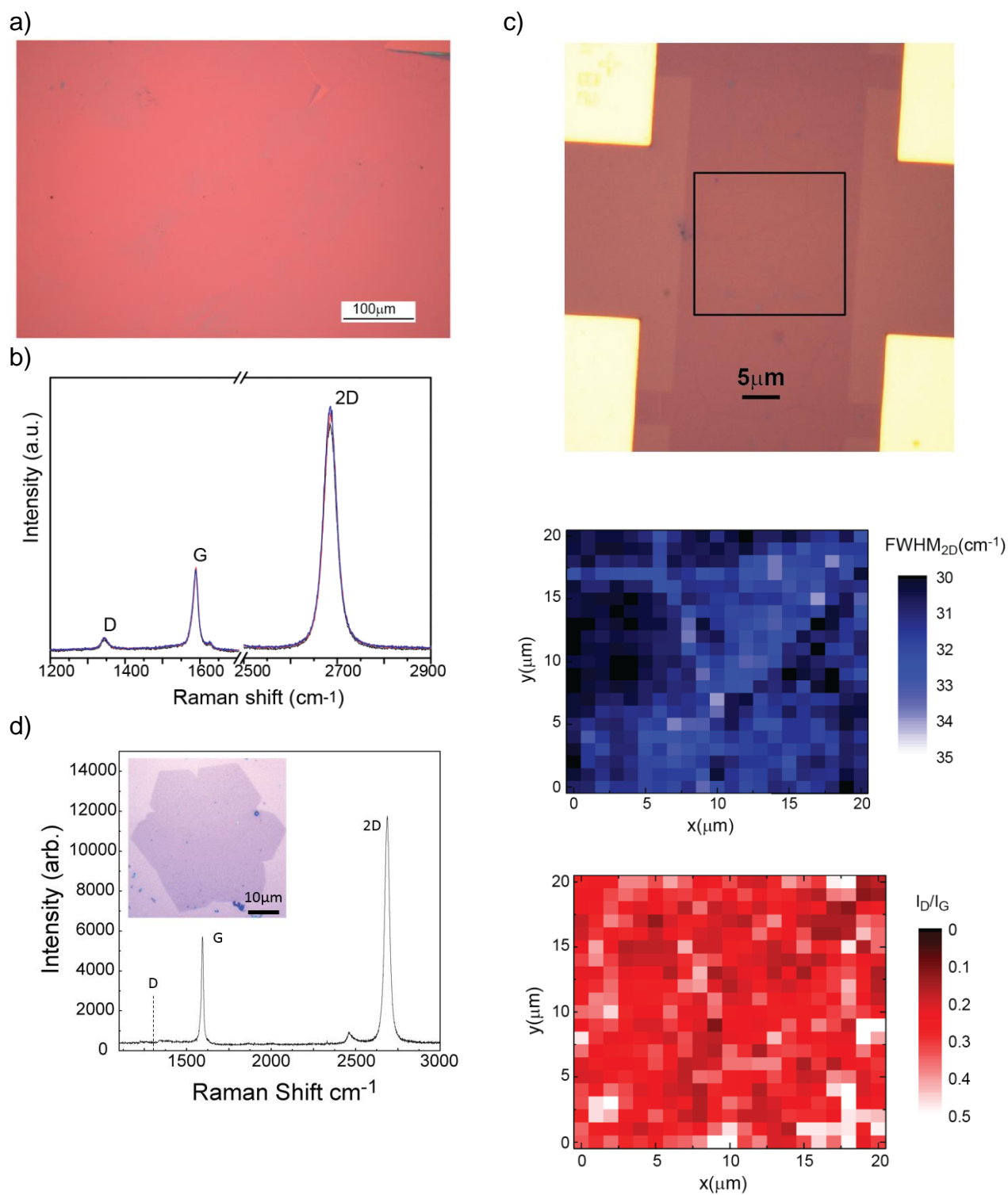

**Figure** S10. a) Optical image of the graphene film transferred on $SiO_2$/Si. The underlying substrate is visible in the upper right corner of the image. b) Three representative Raman spectra plotted for different regions of a continuous graphene grown using the two stage growth technique. c) Top: Optical image of a Hall bar device patterned from the graphene film shown in a). The yellow parts are the Au electrodes. The black square indicates the area used to map the Raman spectra shown in the middle and bottom. Middle: Full width at half maximum (FWHM) of the 2D band. Bottom: The intensity ratio of the D to G peak. d) Raman spectra of a large graphene island taken in the middle of the island. No defect-related D band is observed in this case.

*Touch sensor fabrication and characterization*

The touch sensor device was fabricated using a novel technique where all lithography is performed on the surface of a CVD graphene covered copper foil. Fig. S11 shows the outline of the fabrication process, while Fig. S12 shows images of key processing steps. CVD graphene on copper foils where coated in PMMA and contacts were defined using electron beam lithography, Fig. S11 a and b. The PMMA was developed shown in Fig. S12 a and metallized with 50 nm of gold, Fig. S11c and Fig. S12b. Strips of graphene were made between the contacts by coating the CVD graphene on copper foil with PMMA and defining a mask using electron beam lithography, Fig. S11d. The PMMA was developed and the exposed graphene was etched using $Ar_2/O_2$ reactive ion etching leaving conductive graphene channels between the gold contacts, Fig. S11e and Fig. S12d. The foil was then coated in PMMA again and wet etched in 1 molar $FeCl_3$ solution, Fig. S11f and Fig. S12e. The film was rinsed in ultra-pure water, Fig. S11g, and transferred to a clean PEN substrate, Fig. S11h and Fig. S12f. A PMMA dielectric layer was then coated on the graphene strips transferred to the PEN substrate. A second set of graphene strips were transferred on top of the PMMA/graphene/PEN. The top strips were rotated by 90 degrees with respect to the bottom graphene strips, giving the 2D network of capacitors for the touch sensor.

To characterize the contact and sheet resistance of the graphene films processed in this way we deposited gold contacts without etching graphene strips and transferred the films to a PEN substrate, set out in Fig. S111a-c, f-h.

The two terminal resistance of the graphene strips was measured in air using a probe station and a Keithley source-meter. The capacitance between graphene strips was measured using a Hameg 8118 LCR bridge with 1V AC excitation at 1KHz.

The two terminal resistance was measured as a function the number of squares (distance between probes divided by the sample width) shown in Fig. S12c. The fitted linear gradient is representative of the film resistivity which we estimate to be 1.3KΩ/□, whereas the y intercept of the linear fit is the sum of the contact resistance for the two contacts, estimated to be 68 Ω for each contact.

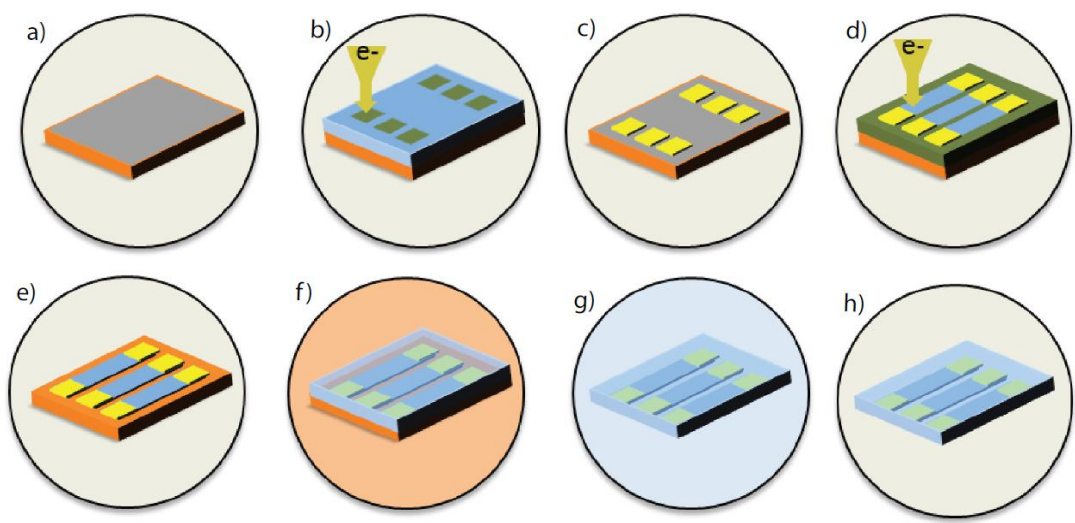

**Figure S11.** The process for fabricating the touch sensor devices. a) Graphene is grown on a copper substrate, b) The foil is coated with PMMA and contacts are exposed using electron beam lithography, c) Exposed regions are developed and metalized with 50nm of gold, d) The foil is coated with PMMA and an etch mask is defined between the gold contacts with electron beam lithography, e) Exposed graphene is etched using an Argon plasma, f)The foil is coated with PMMA and the copper is etched using 1 molar $FeCl_3$, g) The film is washed in ultra-pure water and h) the film is transferred to a PEN substrate.

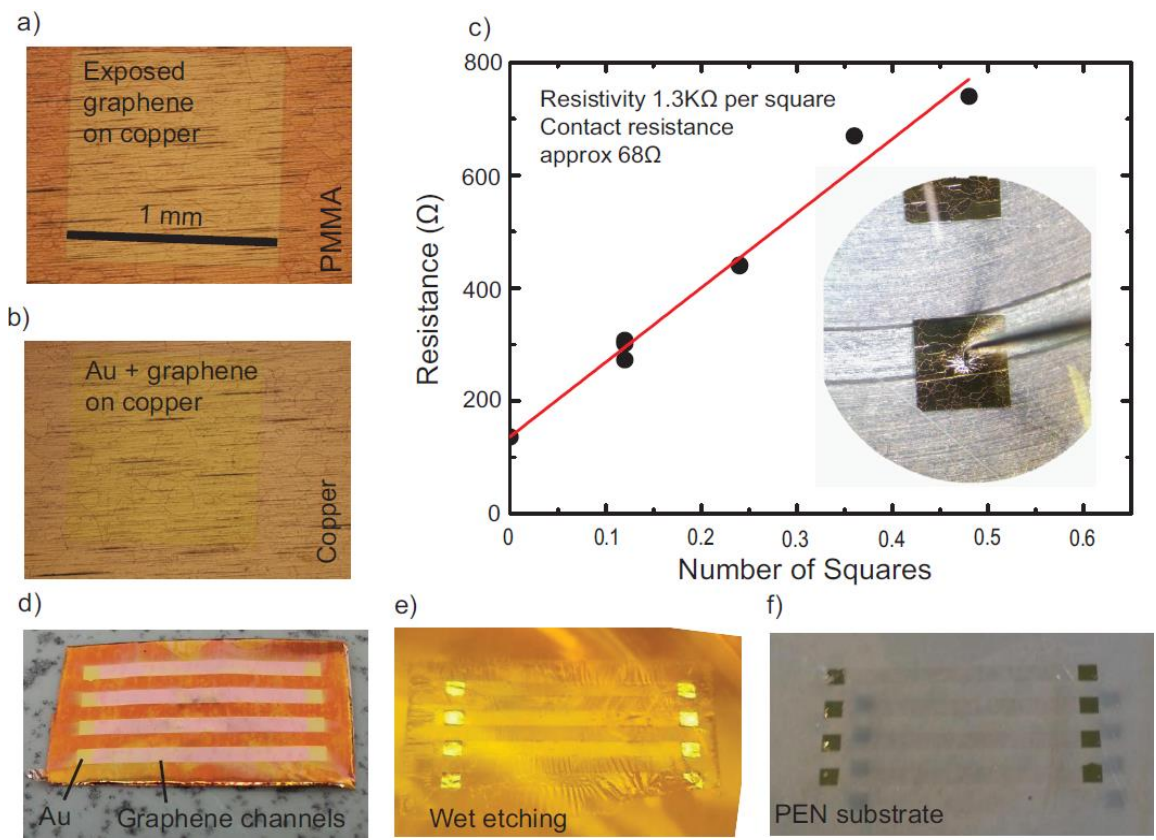

**Figure S12.** a) Shows a window in PMMA after electron beam exposure and development on copper foil coated in CVD graphene, b) Shows a gold square after the metallization gold on top of a copper foil coated with CVD graphene, c) shows the resistance for different separations of gold contacts on graphene transferred to a PEN substrate where the y intercept gives contact resistance and the fitted gradient gives the resistivity, d) Shows gold contacts connected by graphene strips on the surface of the copper foil, e) shows a gold contacts supported by a PMMA film in $FeCl_3$ etchant, f) shows the transferred structure onto a PEN substrate.

*Costing of graphene growth*

The estimation of the cost of graphene growth was performed making several assumptions. There are three main factors affecting the price of producing graphene, the cost of growth gases; the energy cost for achieving the temperatures for growth and the cost of the copper used for growths. These calculations do not consider the cost of growth equipment such as furnaces, flow controllers and quartz tube. The costs were only estimated for published papers that contain enough information to estimate growth cost and the quality area of the graphene.

*Cost of Gases.* The cost of growth gases was estimated by collating the total volume of each gas used from growth times and gas flow rates. The cost per unit volume was then estimated assuming the same price for a set volume of gas [4] allowing for the total cost to be estimated shown in Table S1.

| Gas Type | Cost | Volume (m$^3$) | Cost per Volume £/(m$^3$) |
|---|---|---|---|
| Hydrogen N = 5.5 | £276.12[4] | 8.8 | 31.37 |
| Methane N = 5.5 | £889.97[4] | 10 | 88.99 |
| Argon N = 6 | £339.43[4] | 10.6 | 32.02 |

**Table S1**. The estimation of cost of each different growth gas in £/m$^3$

From each research article, gas flow rates and times were collated shown in Table S2. A typical growth consists of following stages: heating to growth temperature, anneal of the copper foils, graphene island nucleation and the growth stage. Summing the volume of gas used for each stage allowed for the estimation of the total volume of each gas used and the cost of each gas.

| | Hydrogen | | Methane | | Argon | | |
|---|---|---|---|---|---|---|---|
| Article | Volume (m$^3$) | Cost (£) | Volume (m$^3$) | Cost (£) | Volume (m$^3$) | Cost (£) | Total cost (£) |
| Li et. al.[6] (2011) | 2.6×10$^{-4}$ | 0.0081 | 4.5×10$^{-5}$ | 0.0040 | – | – | 0.0121 |
| Venugopal et. al.[14] (2011) | 2.6×10$^{-4}$ | 0.0081 | 4.5×10$^{-5}$ | 0.0040 | – | – | 0.0121 |
| Bae et. al.[7] (2010) | 1.1×10$^{-3}$ | 0.0351 | 1.05×10$^{-3}$ | 0.0934 | – | – | 0.1285 |
| Li et. al.[8] (2011) | 3.2×10$^{-4}$ | 0.0102 | 2.8×10$^{-3}$ | 0.2491 | – | – | 0.2590 |
| Liu et. al.[9] (2011) | 4.3×10$^{-4}$ | 0.0130 | 1.6×10$^{-5}$ | 0.0150 | 1.54×10$^{-2}$ | 0.4940 | 0.5220 |
| Sun et. al.[10] (2012) | 3.3×10$^{-4}$ | 0.0104 | 1.5×10$^{-4}$ | 0.0048* | 1.66×10$^{-2}$ | 0.5334 | 0.5486 |
| Hao et. al[11] (2013) | 9.8×10$^{-3}$ | 0.3059 | 8.3×10$^{-4}$ | 0.0738 | – | – | 0.3797 |
| Chen et. al.[12] (2013) | 3.9×10$^{-3}$ | 0.1220 | 5.85×10$^{-5}$ | 0.1186 | – | – | 0.2406 |
| This work (2015) | 7.5×10$^{-6}$ | 0.0002 | 3.59×10$^{-5}$ | 0.0031 | – | – | 0.0033 |

**Table S2.** The collation of the gas consumption for several graphene growth studies, for the estimation of the total cost of graphene growth gases. The cost for methane has been substituted for that of argon, as argon diluted methane was used.

*Cost of Energy.* To estimate the total energy consumption and cost of each growth process we collated for each different stage of the process, the total growth time, power draw and the cost of electricity in Table S3. The energy consumed during the growth process in a hot wall furnace was estimated assuming an MTI 1200X - 5L tube furnace [5]. The power consumption is assumed to be at maximum during the ramping to the growth temperature (6KW) and that the power consumption scales linearly as a function of temperature to a maximum of 6KW at 1200°C. The energy consumed by plasma based cold walled furnace is estimated at 0.7 KW. The energy consumed by a resistively heated cold walled furnace [this work] is 0.3KW for the ramping to the growth temperature and assumed to scale linearly as a function of temperature to 0.3KW at 1200°C. The cost of electricity is estimated at £0.1352 per KWH [13].

|  | Heating up | | | Anneal and Growth | | |  |
|---|---|---|---|---|---|---|---|
| Article | Time (hours) | Power (KW) | Cost (£) | Time (hours) | Power (KW) | Cost (£) | Total cost (£) |
| Li et. al.[6] (2009) | 1.000 | 6.0 | 0.811 | 0.50 | 5.175 | 0.349 | 1.159 |
| Venugopal et. al.[14] (2011) | 1.000 | 6.0 | 0.811 | 0.50 | 5.175 | 0.349 | 1.159 |
| Bae et. al.[7] (2010) | 0.666 | 6.0 | 0.540 | 1.00 | 5.175 | 0.700 | 1.235 |
| Li et. al.[8] (2011) | 0.666 | 6.0 | 0.540 | 1.50 | 5.175 | 1.049 | 1.590 |
| Liu et. al.[9] (2011) | 0.167 | 6.0 | 0.135 | 0.55 | 5.175 | 0.384 | 0.520 |
| Sun et. al.[10] (2012) | – | – | – | 0.22 | 0.700 | 0.021 | 0.021 |
| Hao et. al[11] (2013) | 0.666 | 6.0 | 0.540 | 15.08 | 5.175 | 10.551 | 11.091 |
| Chen et. al.[12] (2013) | 0.666 | 6.0 | 0.540 | 6.30 | 5.175 | 4.408 | 4.948 |
| This work (2015) | 0.050 | 0.3 | 0.002 | 0.26 | 0.195 | 0.007 | 0.009 |

**Table S3.** The collation energy consumption of each growth procedure, broken down into the heating of the foils to the growth temperature and the growth process.

*Cost of Copper.* We assumed 1cm$^2$ of 25μm thick copper was used in a growth. The cost of copper (99.999 %) is £88.20 for 250cm$^2$ giving a cost for 1cm$^2$ of £0.3528.

*Estimation of total price.* By summing the cost of the growth gases, energy used and the cost of the copper foils we can estimate the total cost of graphene production, shown in Table S4. It is clear from Table S4 that the lowest cost of production is this work and that the limiting factor is the cost of the copper foils used in the growth.

|  | Cost (£) | | | |
|---|---|---|---|---|
| Article | Gas cost | Energy cost | Copper cost | Total cost |
| Li et. al.[6] (2009); Venugopal et. al.[14] (2011) | 0.2590 | 1.159 | 0.3528 | 1.762 |
| Bae et. al.[7] (2009) | 0.1285 | 1.235 | 0.3528 | 1.716 |
| Li et. al.[8] (2011) | 0.0121 | 1.590 | 0.3528 | 1.955 |
| Liu et. al.[9] (2011) | 0.5220 | 0.520 | 0.3528 | 1.35 |
| Sun et. al.[10] (2012) | 0.5486 | 0.021 | 0.3528 | 0.922 |
| Hao et. al[11] (2013) | 0.3797 | 11.091 | 0.3528 | 11.820 |
| Chen et. al.[12] (2013) | 0.2406 | 4.948 | 0.3528 | 5.428 |
| This work (2015) | 0.0033 | 0.009 | 0.3528 | 0.365 |

**Table S4**. The estimation of cost of each price component and the total cost of the growth for each article in £.

## Electronic Quality Factor estimation

To determine the electronic quality factor for each article, the reported field effect mobility was used and the area of the device was estimated from dimensions given or images appearing in the articles. The data was collated into Table S5 and are shown in Figure S11.

| Article | Mobility (cm$^2$/(Vs)) | Area ($\mu$m$^2$) | Electronic Quality Factor (Q) |
|---|---|---|---|
| Li et. al.[6] (2009) | 4050 | 69.66 | $2.82 \times 10^5$ |
| Bae et. al.[7] (2009) | 3175 | 1.16 | $3.68 \times 10^3$ |
| Li et. al.[8] (2011) | 4000 | – | – |
| Liu et. al.[9] (2011) | 3600 | 0.59 | $2.13 \times 10^3$ |
| Venugopal et. al.[14] (2011) | 2450 | 2880 | $7 \times 10^6$ |
| Sun et. al.[10] (2012) | 1800 | 8 | $1.44 \times 10^4$ |
| Hao et. al[11] (2013) | 10800 | 3.9 | $4.21 \times 10^4$ |
| Chen et. al.[12] (2013) | 5200 | 64.9 | $3.33 \times 10^5$ |
| This work (2015) | 2351 | 2500 | $5.87 \times 10^6$ |
| This work (2015) | 1936 | 3750 | $7.2 \times 10^6$ |
| This work (2015) | 1744 | 2500 | $4.36 \times 10^6$ |
| This work (2015) | 3300 | 1250 | $4.13 \times 10^6$ |

**Table S5.** The collation of information required to make the estimation of Electronic Quality Factor for each article. The mobility is the field effect mobility (cm$^2$/Vs), the area is the device area ($\mu$m$^2$) over which the mobility was estimated and the Electronic Quality Factor ($\mu$m$^2$xcm$^2$/(Vs)).

Plotting the price *versus* the mobility shown in Fig. S11a demonstrates the general trend of the cost of production with respect to the mobility. The cost of producing graphene using a cold wall furnace reduces the price significantly when compared to graphene produced in a hot wall furnace while not impacting on the quality of the graphene produced as the general trend would imply.

The quality of cold-wall CVD graphene as compared to that grown with other methods is better assessed using the electronic quality factor (Q) that. As shown in the Figure S11b, graphene grown by resistive heating cold-wall CVD has Q ranging from 4x10$^6$ to 7.2x10$^6$, whereas most reports of monolayer graphene grown by hot-wall CVD have Q ranging from 10$^3$ to 7 x 10$^6$. This demonstrates the enhanced electronic quality range of graphene grown by resistive heating cold-wall CVD over the reported values of monolayer graphene grown by hot-wall CVD. Thus as shown in Figure S11b the cold-wall CVD provides a method to produce high quality graphene at a much lower cost than hot-wall CVD. Employing this method in industry will reduce also the retail price of graphene which currently is as high as 21£/cm² as shown in Figure S11c.

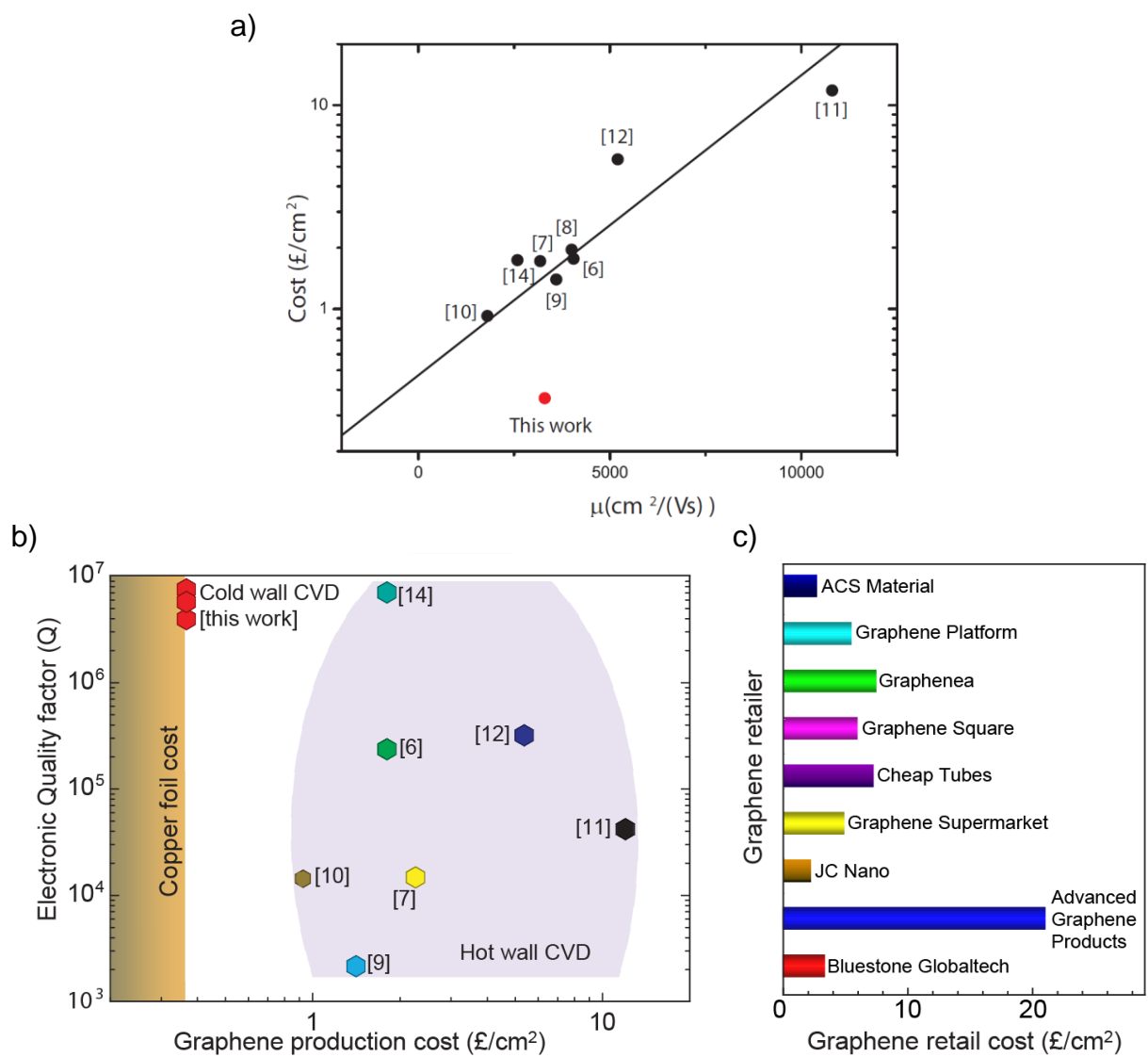

**Figure S11.** a) A plot of the price of graphene production per cm² against the measured mobility. The general trend is a linear fit of the data omitting the data point from this work. b) Estimated cost for different CVD growth processes for monolayer-graphene on Cu plotted against the electronic quality factor, Q. c) Retail cost of monolayer graphene as of April 2015 taken from the website of different suppliers of monolayer graphene grown by CVD on Cu.

*References for the Supporting Information*